\documentclass[aps,prc,twocolumn,superscriptaddress,showpacs,floatfix,nofootinbib]{revtex4-1}
\usepackage{amsmath,graphicx}
\newcommand{\trento}{T$\mathrel{\protect\raisebox{-2.1pt}{R}}$ENTo}

\begin{document}

\title{Reconstructing the impact parameter of proton-nucleus and nucleus-nucleus collisions}

\author{Rudolph Rogly}
\affiliation{
Institut de physique th\'eorique, Universit\'e Paris Saclay, CNRS,
CEA, 91191 Gif-sur-Yvette, France}
\affiliation{MINES ParisTech, PSL Research University, 60 Boulevard Saint-Michel, 75006 Paris, France}
\author{Giuliano Giacalone}
\affiliation{
Institut de physique th\'eorique, Universit\'e Paris Saclay, CNRS,
CEA, 91191 Gif-sur-Yvette, France}  
\author{Jean-Yves Ollitrault}
\affiliation{
Institut de physique th\'eorique, Universit\'e Paris Saclay, CNRS,
CEA, 91191 Gif-sur-Yvette, France} 

\begin{abstract}
In proton-nucleus and nucleus-nucleus collision experiments, one determines the centrality of a collision according to the multiplicity or energy deposited in a detector. This serves as a proxy for the true collision centrality, as defined by the impact parameter. We show that the probability distribution of impact parameter in a given bin of experiment-defined centrality can be reconstructed without assuming any specific model for the collision dynamics, in both proton-nucleus and nucleus-nucleus systems. The reconstruction is reliable up to about 10\% centrality, and is more accurate for nucleus-nucleus collisions. We perform an application of our procedure to experimental data from all the CERN Large Hadron Collider (LHC) collaborations, from which we extract, in Pb+Pb and $p$+Pb collisions, the corresponding distributions of impact parameter.
\end{abstract}

\maketitle

\section{Introduction}
The impact parameter of a heavy-ion collision, $b$, is not a directly measurable quantity. 
In experiments, the \textit{centrality} of a collision has to be inferred from the amount of hits~\cite{Back:2000gw,Adler:2001yq,Adler:2004zn,Abelev:2013qoq,Adam:2014qja,Adamczewski-Musch:2017sdk}, or energy~\cite{Chatrchyan:2011pb,ATLAS:2011ah,Aad:2015zza} a certain collision produces in a specific detector. Its relation with the impact parameter is not one-to-one. A given value of experiment-defined centrality corresponds to a distribution of impact parameters. 
In this paper, we show that the probability distribution of impact parameter at a given centrality can be reconstructed to a good approximation without assuming any specific model of the collision, in particular, without resorting to the concept of participant nucleons.

In a recent publication, we have carried out such a reconstruction for the distribution of impact parameter of central nucleus-nucleus collisions~\cite{Das:2017ned}. Here, we present an improvement of the reconstruction procedure. As we shall see in the following, this enables us to infer the distribution of $b$ as well in proton-nucleus systems, which were beyond the applicability of the previous framework.

Let us briefly describe the improvement. In Ref.~\cite{Das:2017ned}, we denoted by $n$ the quantity which is used to define the centrality experimentally~\cite{Broniowski:2001ei}, and 
we assumed that the fluctuations of $n$ were Gaussian for a fixed impact parameter. The mean and width of the Gaussian were then inferred by fitting the measured distribution of $n$. 
In this paper, the Gaussian kernel is replaced with a gamma distribution. As explained in Sec.~\ref{s:kernel}, both distributions have the same number of parameters, but the gamma distribution presents two dramatic advantages: $i$) it  satisfies the requirement $n\ge 0$; $ii$) in the regime of large $n$, it falls asymptotically much more mildly than a Gaussian ($e^{-n}$ instead of $e^{-n^2}$), and this radically improves the description of multiplicity fluctuations in proton-nucleus collisions, which present a very long tail. 

Apart from this modification of the fluctuation kernel, the reconstruction goes along the same lines as in Ref.~\cite{Das:2017ned}.
We validate the procedure in Sec.~\ref{s:validation} using a specific model of the collision dynamics where the impact parameter is known in each event, so that we are able to assess the accuracy of the reconstruction method. 
In Sec.~\ref{s:data}, the procedure is then applied to Large Hadron Collider (LHC) data on Pb+Pb and $p$+Pb collisions.

\section{Fluctuation kernel}
\label{s:kernel}

Current experimental analyses define the centrality of a heavy-ion collision according to a single observable, $n$, which is typically the number of hits or the energy measured in a dedicated calorimeter. This quantity is always positive, and extensive, in the sense that in one event it is given by the sum of many contributions collected in the detector.
$n$ does not have a one-to-one correspondence with the impact parameter, and its relation with $b$ is inferred by matching the measured distribution $P(n)$ to the distribution provided by some Monte Carlo model of the collision dynamics, typically, the Glauber Monte Carlo model~\cite{Loizides:2014vua}.
In the current state-of-the-art, such procedure relies entirely on the concept of participant nucleons~\cite{Miller:2007ri}, as $n$ in the simulations always depends on either the number of nucleons that participate in the collision of two nuclei in a nucleus-nucleus collision, or the number of nucleons hit by a projectile proton in a proton-nucleus collision.
Here we do something simpler. Our goal is to infer information about $b$  using the measured distributions $P(n)$ without resorting to any simulation of the whole collision process. 

The only ingredient we need to model is the probability distribution of $n$ for fixed $b$, $P(n|b)$. The Gaussian distribution, which was our choice in Ref.~\cite{Das:2017ned}, is the most natural candidate, but turns out to be technically wrong because it has a left tail which extends to negative values of $n$. This is unimportant in practice as long as the standard deviation is much smaller than the mean, because almost all probability is on the positive side. When studying systems with large fluctuations, however, the positivity issue can no longer be ignored, and one should use a distribution where {\it all\/} the probability is on the positive half-line.

To overcome this issue, we change our fluctuation kernel into a gamma distribution, a non-negative distribution whose density reads:
\begin{equation}
\label{Gamma}
P(n|b)=\frac{1}{\Gamma(k)\theta^k}n^{k-1} e^{-n/\theta}.
\end{equation}
$k$ and $\theta$ are two positive parameters, which generally depend on $b$. $k$ is dimensionless, while $\theta$ has the same dimension as the observable $n$. The gamma distribution is normalized on the positive half-line: $\int_0^{+\infty}P(n|b)dn=1$. 
The mean and the standard deviation are given by:
\begin{eqnarray}
  \label{meanvariance}
\bar  n &=&k\theta\cr
\sigma_n &=&\sqrt{k}\theta.  
\end{eqnarray} 
The gamma distribution has the following additivity property: if $n$ is the sum of $N$ independent random variables $n_i$, $i=1,\cdots,N$, where each $n_i$ is distributed according to a gamma distribution with parameters $k_i$ and a common value of $\theta$, then $n$ also follows a gamma distribution, with parameters $k=\sum_i k_i$ and $\theta$. This property makes the gamma distribution an attractive choice when $n$ is the multiplicity or energy produced in a collision, as $n$ is usually modeled as the sum of several independent  contributions, e.g., participant nucleons~\cite{Miller:2007ri} or wounded quarks~\cite{Eremin:2003qn,Bialas:2007eg,Bozek:2016kpf}. Because of this additivity property, $k$ can be viewed as an extensive parameter, which is proportional to the system size, and $\theta$ as an intensive parameter, independent of the system size. 

The gamma distribution coincides with a Gaussian in the limit $k\gg 1$. This is precisely the limit of small fluctuations, where the Gaussian distribution is universal by virtue of the central limit theorem. Therefore, we expect the gamma and the Gaussian fluctuation kernels to provide identical results when the central limit theorem applies, and the gamma kernel to be a viable choice when Gaussianity is spoiled by large fluctuations. 

The gamma distribution is also a natural choice in the context of high-energy physics because it is essentially a continuous version of the negative binomial distribution (NBD)~\cite{Bozek:2013uha}, with which it coincides when the mean value is much larger than unity.\footnote{If one fixes the 2 parameters of the gamma distribution in such a way that the mean, $\bar n$, and the variance, $\bar n\theta$, are equal to those of the NBD distribution, then $\theta>1$, because the variance of the NBD is always larger than the mean.
The equivalence between the two distributions in the limit $\bar n\gg 1$ can be seen by comparing the higher-order cumulants. 
The standardized skewness is $2\sqrt{\bar n/\theta}$ for the gamma distribution, and $2\sqrt{\bar n/\theta}-2/\sqrt{\bar n\theta}$ for the NBD, hence the difference is $2/\sqrt{\bar n\theta}\ll 1$. 
The excess kurtosis is $6\theta/\bar n$ for the gamma distribution, and $6\theta/\bar n-6/\bar n+1/(\bar n\theta)$ for the NBD, so the difference is negligible in the limit $\bar n\gg 1$.} 
The NBD has long been used to fit multiplicity distributions in high-energy collisions~\cite{Giovannini:1985mz}.
Most model calculations of multiplicity distributions in proton-nucleus collisions~\cite{Tribedy:2011aa,Dumitru:2012yr,Moreland:2012qw} involve NBD fluctuations, which are also expected in the color-glass condensate picture of high-energy QCD~\cite{Gelis:2009wh}. 
Further, typical Monte Carlo models used to simulate the initial state of nucleus-nucleus and proton-nucleus collisions implement such fluctuations using the gamma distribution~\cite{Broniowski:2007nz,Moreland:2014oya}.

\begin{figure*}[t!]
\begin{center}
\includegraphics[width=\linewidth]{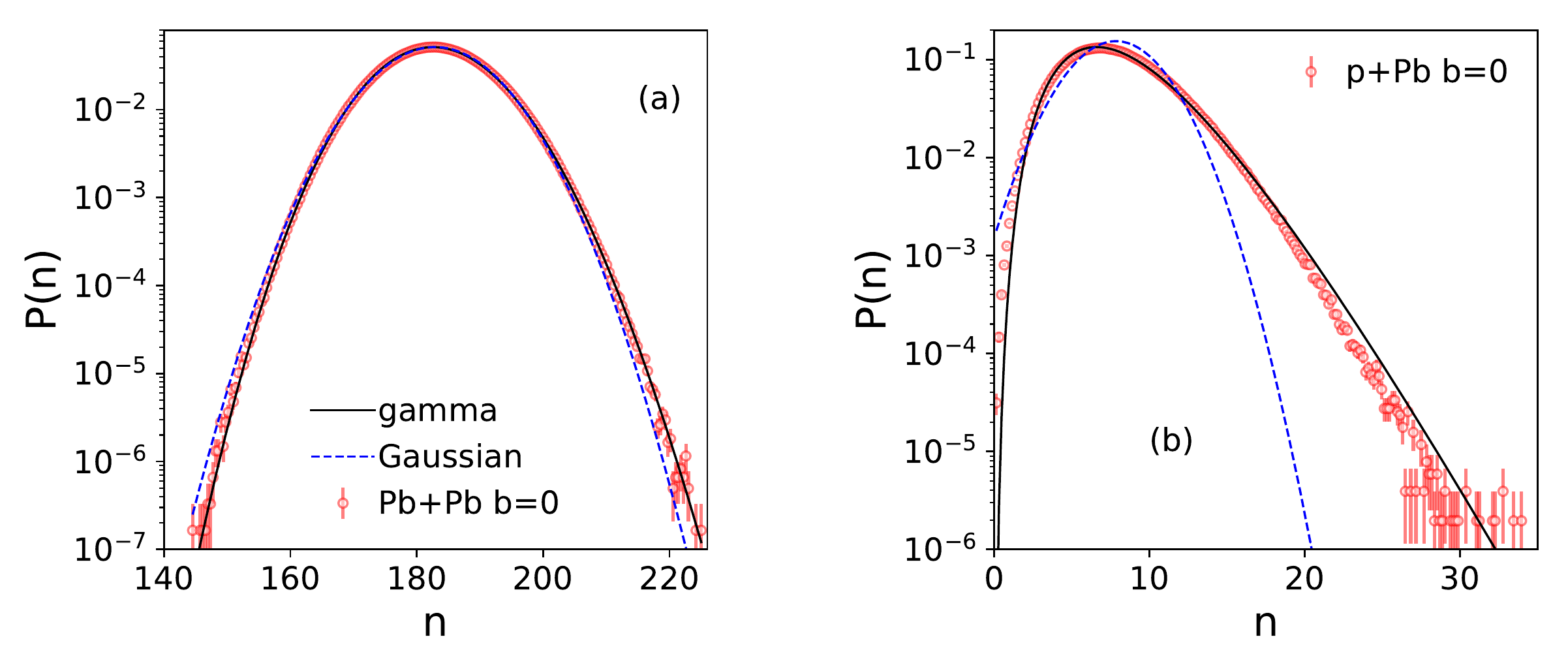} 
\end{center}
\caption{(Color online) 
\label{fig:kernel}
Probability distribution of the total entropy $n$ in the {\trento} model for: (a) Pb+Pb collisions at $b=0$; (b) $p$+Pb collisions at $b=0$. The energy is $\sqrt{s_{\rm NN}}=5.02$~TeV in both cases. Lines are 2-parameter fits using a Gaussian (dashed lines) or a gamma distribution (full lines). 
}
\end{figure*}

We illustrate the shape of $P(n|b)$ by simulating Pb+Pb and $p$+Pb collisions at fixed impact parameter using the \trento{} model of initial conditions \cite{Moreland:2014oya}.
 There are essentially two free parameters in {\trento}, a parameter $p$ which specifies the respective contributions of projectile and target participants, and a parameter $k$ which tunes the magnitude of event-to-event fluctuations.\footnote{The fluctuation parameter $k$ of \trento{} should not be confused with the parameter $k$ of Eq.~(\ref{Gamma}).}
 For Pb+Pb collisions, we use the phenomenologically successful parametrization $p=0$ and $k=1.6$~\cite{Bernhard:2016tnd}, whereas for $p$+Pb collisions, we use $p=1$, corresponding to a Glauber Monte Carlo prescription, and $k=0.6$. 
We have checked that this setup for $p$+Pb collisions provides a distribution $P(n)$ which matches to a good extent the distributions $P(E_T)$ measured by the ATLAS~\cite{Aad:2015zza} collaboration, and the distribution $P(N_{trk})$ measured by the CMS collaboration \cite{Chatrchyan:2013nka}.
We use these simulations merely as an illustration, therefore, it is irrelevant whether the chosen parametrizations give a precise description of data or not.
We generate collisions at zero impact parameter: $1.5\times 10^7$ Pb+Pb collisions and $3\times 10^6$ $p$+Pb collisions.
The distributions of $n$ are displayed in Fig.~\ref{fig:kernel}.

Figure~\ref{fig:kernel}(a) presents the probability distribution $P(n)$ in Pb+Pb collisions. The standard deviation of the distribution is $\sigma_n\sim 8$, much smaller than the mean $\bar n\sim 182$. In this regime of small fluctuations, one expects $P(n)$ to be well described either by a gamma or a Gaussian distribution. 
The corresponding fits are shown as lines in the figure. The fit with a gamma distribution works better, and the reason is that the gamma naturally captures the skewness of $P(n)$ (its skewness is larger than the actual value by 15\%). The value of $k$ in Eq.~(\ref{Gamma}) is $k=553$, which satisfies the condition $k\gg 1$, and the regime of small fluctuations. Interestingly then, this result implies that the gamma kernel may bring a significant improvement over the Gaussian one even in large Pb+Pb systems.

We move on now to proton-nucleus collisions. 
Figure~\ref{fig:kernel}(b) displays $P(n|b=0)$ for $p$+Pb collisions. Note that this distribution is much broader than the previous one.
In the model, this is due to the fact that we are far away from the regime of validity of the central limit theorem, because of the lower number of participant nucleons.
For this reason, $P(n)$ falls off exponentially at large $n$, with a very long tail. 
While this exponential decrease at large $n$ is essentially put by hand in this particular model, it is predicted more generally by models of high-energy QCD scattering~\cite{Liou:2016mfr}. 
As expected, the Gaussian parametrization is not viable in this situation.
The gamma distribution, on the other hand, provides a good fit. As for the Pb+Pb collisions, it naturally reproduces the skewness of the distribution, which is overestimated by 20\%. 

It may seem a tautology that the gamma distribution provides a good fit to this model calculation, since fluctuations in the {\trento} model are implemented using the gamma distribution. 
More precisely, for proton-nucleus collisions, $P(n)$ would be exactly a gamma distribution if the number of participants, $N_{\rm part}$, were fixed.
However, $N_{\rm part}$ fluctuates significantly at a fixed impact parameter, and this explains why the fit in Fig.~\ref{fig:kernel}(b) is not perfect, although very reasonable. 
Our point is that the gamma distribution is a natural parametrization, irrespective of any specific model, for the fluctuations of a positive extensive quantity. 

\section{Method and validation}
\label{s:validation}

\begin{figure*}[t!]
\begin{center}
\includegraphics[width=\linewidth]{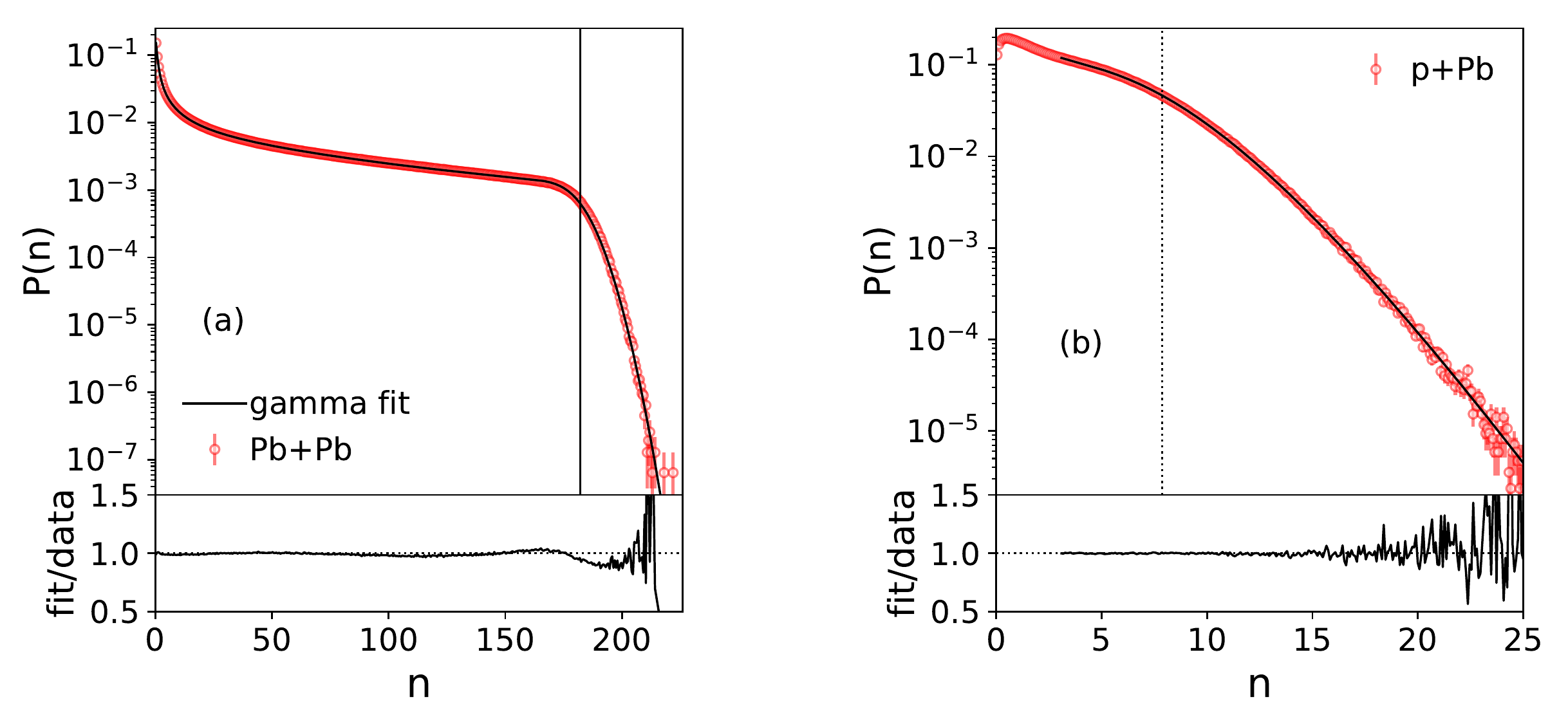} 
\end{center}
\caption{(Color online) 
\label{fig:2}
Probability distribution of the total entropy $n$ in the {\trento} model for: (a) $2.8\times 10^7$ Pb+Pb collisions; (b) $10^7$ $p$+Pb collisions. 
The only difference with Fig.~\ref{fig:kernel} is that there is no restriction on impact parameter. 
Lines are fits using Eqs.~(\ref{Gamma}), (\ref{cbintegral}) and (\ref{paramk}) (see text).
In each panel, vertical lines indicate the position of the knee of the histogram (see text) returned by the fit (solid lines) or calculated directly (dotted lines). 
}
\end{figure*}

We now describe how the probability distribution of impact parameter at a given centrality can be reconstructed using experimental data, where by experimental data we mean the distribution of $n$, $P(n)$, integrated over all impact parameters. The method is identical to that of Ref.~\cite{Das:2017ned}, except that we replace the Gaussian fluctuation kernel with a gamma distribution. In Fig.~\ref{fig:2} we show $P(n)$ from model calculations of Pb+Pb and $p$+Pb collisions, using the same \trento{} setups as in Fig.~\ref{fig:kernel}. The values of $n$ in the tail of the distribution correspond essentially to those in Fig.~\ref{fig:kernel}, that is, to central collisions, while smaller values of $n$ are produced by collisions at larger impact parameters.

We first derive the expression of $P(n)$ using the model introduced in Sec.~\ref{s:kernel}. We decompose $P(n)$ into contributions coming from all impact parameters: 
\begin{equation}
\label{pnb1}
P(n)=\int_0^{\infty} P(n,b)db, 
\end{equation}
where $P(n,b)$ denotes the joint probability of $n$ and $b$, which is not known in an experiment. 
This joint probability can be decomposed as $P(n,b)=P(n|b)P(b)$, where $P(b)$ is the probability distribution of impact parameter, and 
$P(n|b)$ is the distribution of $n$ at fixed $b$. 
Therefore, we rewrite Eq.~(\ref{pnb1}) as:
\begin{equation}
\label{pnb}
  P(n)=\int_0^{\infty} P(n|b)P(b)db, 
\end{equation}
and we take $P(n|b)$ from Eq.~(\ref{Gamma}).
The probability distribution of $b$ reads
\begin{equation}
  \label{probab}
  P(b)=\frac{1}{\sigma_{\rm inel}}2\pi bP_{\rm inel}(b),
\end{equation}
where $\sigma_{\rm inel}$ is the inelastic nucleus-nucleus or proton-nucleus cross section, depending on the system under study, and $P_{\rm inel}(b)$ is the probability for an inelastic collision to occur at impact parameter $b$, which is close to 1 except for peripheral collisions.

The idea is to find two functions $k(b)$ and $\theta(b)$ to use in Eq.~(\ref{Gamma}) so that $P(n)$ defined by Eq.~(\ref{pnb})  matches the observed $P(n)$. As pointed out in Ref.~\cite{Das:2017ned}, this problem is underconstrained, in the following sense: one cannot extract two unknown functions $k(b)$ and $\theta(b)$ (or equivalently, the mean and standard deviation $\bar n(b)$ and $\sigma_n(b)$) from a single distribution $P(n)$.
More precisely, the information about the width $\sigma_n$ lies in the tail of $P(n)$, which corresponds to $b\sim 0$. 
The rest of the distribution, corresponding to smaller values of $n$, gets contributions from several impact parameters, so that fluctuations are averaged over and $P(n)$ only contains information about the mean $\bar n(b)$~\cite{Broniowski:2001ei}. 
Therefore, one cannot infer from data alone how $\sigma_n(b)$ varies with $b$.

Throughout this paper, we assume for simplicity that the variance is proportional to the mean, that is, $\sigma_n(b)^2/\bar n(b)$ is independent of $b$.
This holds for a superposition of $N$ independent sources with the same distribution, in the sense that the ratio is independent of $N$. 
As shown in Ref.~\cite{Das:2017ned}, final results are robust with respect to this hypothesis, as long as the study is restricted to fairly central collisions (typically 0-10\% in Pb+Pb collisions). 
Eq.~(\ref{meanvariance}) gives $(\sigma_n)^2/\bar n=\theta$. Hence we assume that $\theta$ is independent of $b$. 
The physical explanation is that $\theta$ is an intensive parameter, and should therefore depend weakly on the system size.

On the other hand, the parameter $k$ in Eq.~(\ref{Gamma}) is allowed to depend on $b$. It can be any smooth positive, monotonically decreasing function.\footnote{The mean of the distribution in Eq.~(\ref{Gamma}) is $k\theta$, and larger impact parameters correspond to smaller $n$, hence $k$ should decrease with impact parameter.} As in Ref.~\cite{Das:2017ned}, we change variables from $b$ to its cumulative probability distribution $c_b$:
\begin{equation}
  \label{defcb}
  c_b=\int_0^b P(b')db'. 
\end{equation}
$c_b$ is the \textit{true} centrality, defined according to impact parameter, and it was dubbed $b$-centrality in Ref.~\cite{Das:2017ned}.
The advantage of this change of variable is that the integral over impact parameter, Eq.~(\ref{pnb}), simplifies to
\begin{equation}
\label{cbintegral}
  P(n)=\int_0^1 P(n|c_b) dc_b, 
  \end{equation}
where $P(n|c_b)=P(n|b)$ denotes the probability distribution of $n$ at fixed $c_b$, i.e., fixed $b$. 

Now, we parametrize the variation of $k$ with $c_b$ using
\begin{equation}
  \label{paramk}
k(c_b)=k_{\rm max} \exp\left(-\sum_{j=1}^{J} a_j (c_b)^j\right), 
  \end{equation}
where the exponential guarantees positivity. 
This value is then inserted into the gamma distribution, Eq.~(\ref{Gamma}):
\begin{equation}
  \label{Gamma2}
P(n|c_b)=\frac{1}{\Gamma(k(c_b))\theta^k}n^{k(c_b)-1} e^{-n/\theta}.
\end{equation}
We fit $P(n)$ to data using Eqs.~(\ref{cbintegral}) and (\ref{Gamma2}). 
The fit parameters are $\theta$, $k_{\rm max}$ and the coefficients $a_j$. One can increase the degree of the polynomial until the fit is perfect. 

\begin{figure*}[t!]
\begin{center}
\includegraphics[width=\linewidth]{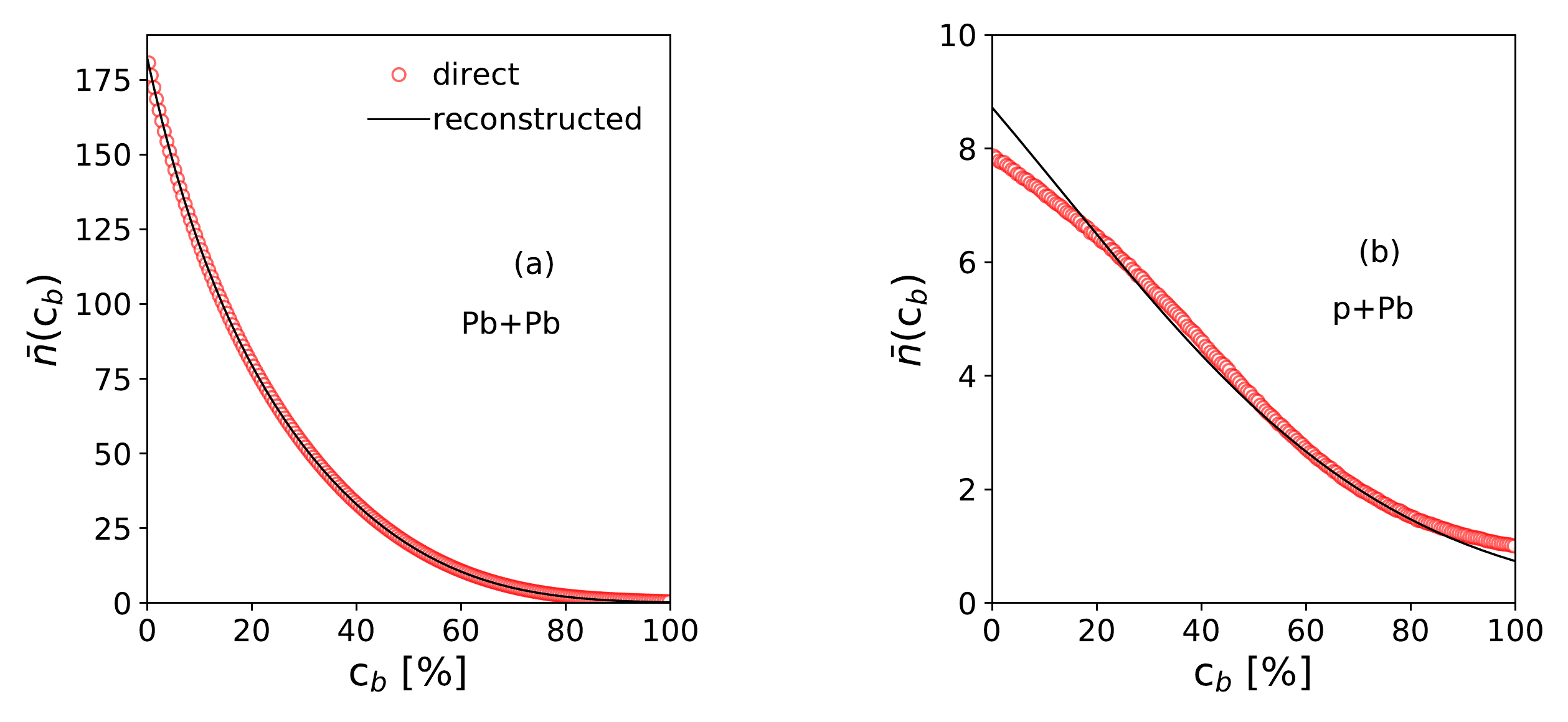} 
\end{center}
\caption{(Color online) 
\label{fig:3}
Mean value of $n$ as a function of $c_b$ returned by the fit (lines) and calculated directly by binning in $c_b$ (symbols), for the same {\trento} events as in Fig.~\ref{fig:2}. (a) Pb+Pb collisions. (b) $p$+Pb collisions. 
}
\end{figure*}

We now test the procedure on simulated data, by fitting the Monte Carlo results shown in Fig.~\ref{fig:2}. Fits are shown as lines. For Pb+Pb collisions, we use a polynomial of degree $J=4$ in Eq.~(\ref{paramk}), so that there are 6 fit parameters. The fit is excellent, across the full range of $n$. It is essentially unchanged if we replace the gamma distribution by a Gaussian distribution, as done in Ref.~\cite{Das:2017ned}.\footnote{Therefore, in this figure and the following figures, we do not show anymore results obtained using a Gaussian kernel, which are essentially identical.} The advantage of the gamma kernel is that one can fit the whole distribution, without the need for a cutoff at small $n$, as required by Gaussian fluctuations in order to minimize the negative tail~\cite{Das:2017ned}. 
For $p$+Pb collisions, we restrict the fit to the rightmost half of the histogram (0-50\% centrality), for reasons explained below, and we use a polynomial of degree $J=2$ in Eq.~(\ref{paramk}), so that there are 4 fit parameters.
The fit is also of excellent quality.


The fit returns the parameters of the gamma distribution in Eq.~(\ref{Gamma2}), i.e., it reconstructs the probability of $n$ at fixed impact parameter. 
In particular, the mean value of $n$ at fixed $c_b$ is given by Eq.~(\ref{meanvariance}):  $\bar n(c_b)=k(c_b)\theta$. In the model calculation, we can also calculate this quantity directly by binning events in $c_b$ and evaluating the mean of $n$ in each bin. The comparison between the fit result and the direct calculation is displayed in Fig.~\ref{fig:3}. The reconstruction is excellent for Pb+Pb collisions, as expected from Ref.~\cite{Das:2017ned}. For $p$+Pb collisions, the reconstruction is less accurate. 
A good indicator of the accuracy of the reconstruction is the location of the \textit{knee} of the distribution, defined as the mean value of $n$ for $b=0$: $n_{\rm knee}\equiv\bar n(c_b=0)$~\cite{Das:2017ned}. 
Its position is shown by vertical lines in Fig.~\ref{fig:2}.
It is overestimated by 11\% in $p$+Pb collisions, while it is reconstructed with better than 1\% accuracy in Pb+Pb collisions.
The standard deviation $\sigma_n$ for central collisions, $\sigma_n(0)=\sqrt{k(0)}\theta$, is overestimated by 4\% in Pb+Pb collisions and underestimated by 5\% in $p$+Pb collisions. 

We now explain why we only fit the 0-50\% centrality range in $p$+Pb collisions.
The idea between our procedure is that $P(n)$ consists of a body and a tail, where the tail contains the information about $\sigma_n(0)$, while the body contains the information about $\bar n(c_b)$~\cite{Broniowski:2001ei,Das:2017ned}.
The latter statement is true only if fluctuations are relatively small, $\sigma_n\ll \bar n$~\cite{Broniowski:2001ei}.
This does not hold for peripheral collisions: 
$P(n)$ for small $n$ depends not only on $\bar n(c_b)$, but also on the width $\sigma_n(c_b)$.
Since a simultaneous fit of $\bar n(c_b)$ and $\sigma_n(c_b)$ using $P(n)$ would be underconstrained, we eliminate peripheral collisions from the fit by introducing a cutoff.
We have checked that the reconstructed value of $n_{\rm knee}$ 
varies mildly, approximately by 3\%, as we vary the centrality cutoff from 30\% up to 60\%, beyond which the quality of the 4-parameter fit decreases rapidly.
Therefore, we conclude that the reconstruction of $n_{knee}$ is robust in $p$+Pb collisions.
Regarding Pb+Pb collisions, we find that excellent fits are obtained even without eliminating peripheral collisions. We will keep a small cutoff at small $n$ only in the analysis of experimental Pb+Pb data in Sec.~\ref{s:data}. 


\begin{figure*}[t!]
\begin{center}
\includegraphics[width=\linewidth]{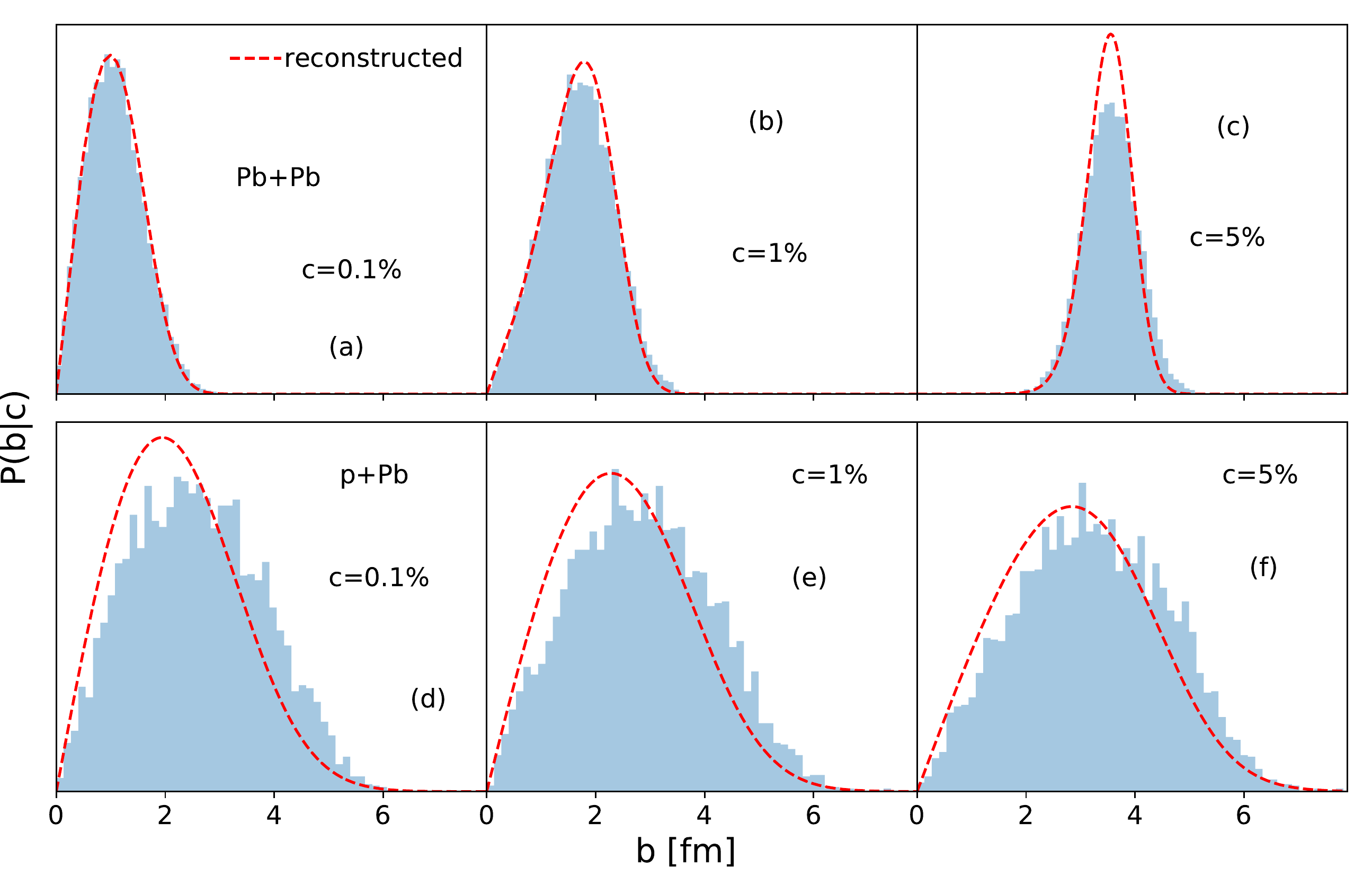} 
\end{center}
\caption{(Color online) 
\label{fig:4}
Probability distribution of $b$ at fixed centrality returned by the fit (lines) and calculated directly by binning in $b$ (shaded area), for the same {\trento} events as in Fig.~\ref{fig:2}. (a)--(c): Pb+Pb collisions. (d)--(f): $p$+Pb collisions.  (a),(d): $c=0.1\%$. (b),(e): $c=1\%$. (c),(f): $c=5\%$. 
}
\end{figure*} 
Once the probability of $n$ at fixed $c_b$ is reconstructed, the probability distribution of impact parameter, $b$, at fixed $n$ is given immediately by Bayes' theorem:
\begin{equation}
  \label{bayes}
P(b|n)=\frac{P(n|b)P(b)}{P(n)}, 
\end{equation}
where $P(n|b)=P(n|c_b)$ is given by Eq.~(\ref{Gamma2}) upon changing variables from $c_b$ to $b$ using $c_b\simeq\pi b^2/\sigma_{\rm inel}$\footnote{The value of $\sigma_{\rm inel}$ is calculated in the \trento{} model. We find $\sigma_{\rm inel}=777$~fm$^2$ in Pb+Pb collisions, and $\sigma_{\rm inel}=210$~fm$^2$ in p+Pb collisions}, and $P(b)$ is given by Eq.~(\ref{probab}), with the approximation $P_{\rm inel}(b)\simeq 1$.
We apply Eq.~(\ref{bayes}) to the \trento{} simulation.
We compare the results thus obtained from the fitting procedure with a direct calculation in which we select events in a narrow interval of $n$,\footnote{Specifically, the width of the interval is 0.05\% in centrality, where centrality is defined in Eq.~(\ref{defc}).} and compute the distribution of $b$ for these events.
The comparison is displayed in Fig.~\ref{fig:4}. 
The curves are labeled by the centrality fraction $c$, defined as the cumulative probability distribution of $n$:
\begin{equation}
  \label{defc}
c\equiv \int_n^{\infty} P(n')dn'. 
\end{equation}
The largest values of $n$ correspond by definition to the most central collisions, i.e., in heavy-ion terminology, the smallest values of $c$. 

Results for Pb+Pb collisions are shown in panels (a)--(c) of Fig.~\ref{fig:4} for $c=$~0.1, 1, 5\%. The distributions are shifted to larger values of $b$ as the centrality $c$ increases, as expected. The agreement between the reconstructed and the direct distributions is excellent for the most central collisions. A small difference with the direct calculation appears at $c=5\%$, confirming that our reconstruction does not capture the centrality dependence of the fluctuations, at least in this particular model~\cite{Das:2017ned}. 

Results for $p$+Pb collisions are shown in panels (d) to (f). 
The impact parameter distribution is much broader than in Pb+Pb collisions, due to the larger fluctuations, and its dependence on centrality is also much milder.
The reconstruction is less perfect but still very reasonable.
The reconstructed distribution is narrower than the true distribution. 
In Ref.~\cite{Das:2017ned}, we pointed out that the width of the distribution in central collisions is determined by a single quantity, which is the fraction of events on the right of the knee of the distribution, that is, the centrality of the knee, $c_{\rm knee}$. The smaller $c_{\rm knee}$, the narrower the distribution. 
The direct calculation gives $c_{\rm knee}=12.5\%$, while the reconstructed value is $c_{\rm knee}=8.7\%$, roughly 30\% too low. This error explains the modest discrepancy between the direct and the reconstructed distributions. 
By contrast, in Pb+Pb collisions, $c_{\rm knee}$ is reconstructed to better than $5\%$ relative accuracy, which explains why the reconstructions in panels (a) and (b) are essentially perfect. 

\section{Application to data} 
\label{s:data}

\begin{table}[b!]
\begin{tabular}{|c|r|r|}
\hline
&Pb+Pb&$p$+Pb\cr
\hline
ALICE &V0A~\cite{Abelev:2013qoq}& V0A~\cite{Adam:2014qja}\cr
ATLAS &FCal $E_T$~\cite{ATLAS:2011ah} & FCal $E_T$~\cite{Aad:2014lta}\cr
CMS &HF $E_T$~\cite{Chatrchyan:2011pb}&$N_{\rm trk}^{\rm offline}$~\cite{Chatrchyan:2013nka}\cr
LHCb &&$N_{\rm VELO}^{\rm hit}$~\cite{Aaij:2015qcq}\cr
\hline
\end{tabular}
\caption{\label{tabledata} 
List of observables used by the LHC collaborations to define the collision centrality, along with references where their distributions are published. 
}
\end{table}
\begin{figure*}[t!]
\begin{center}
\includegraphics[width=.95\linewidth]{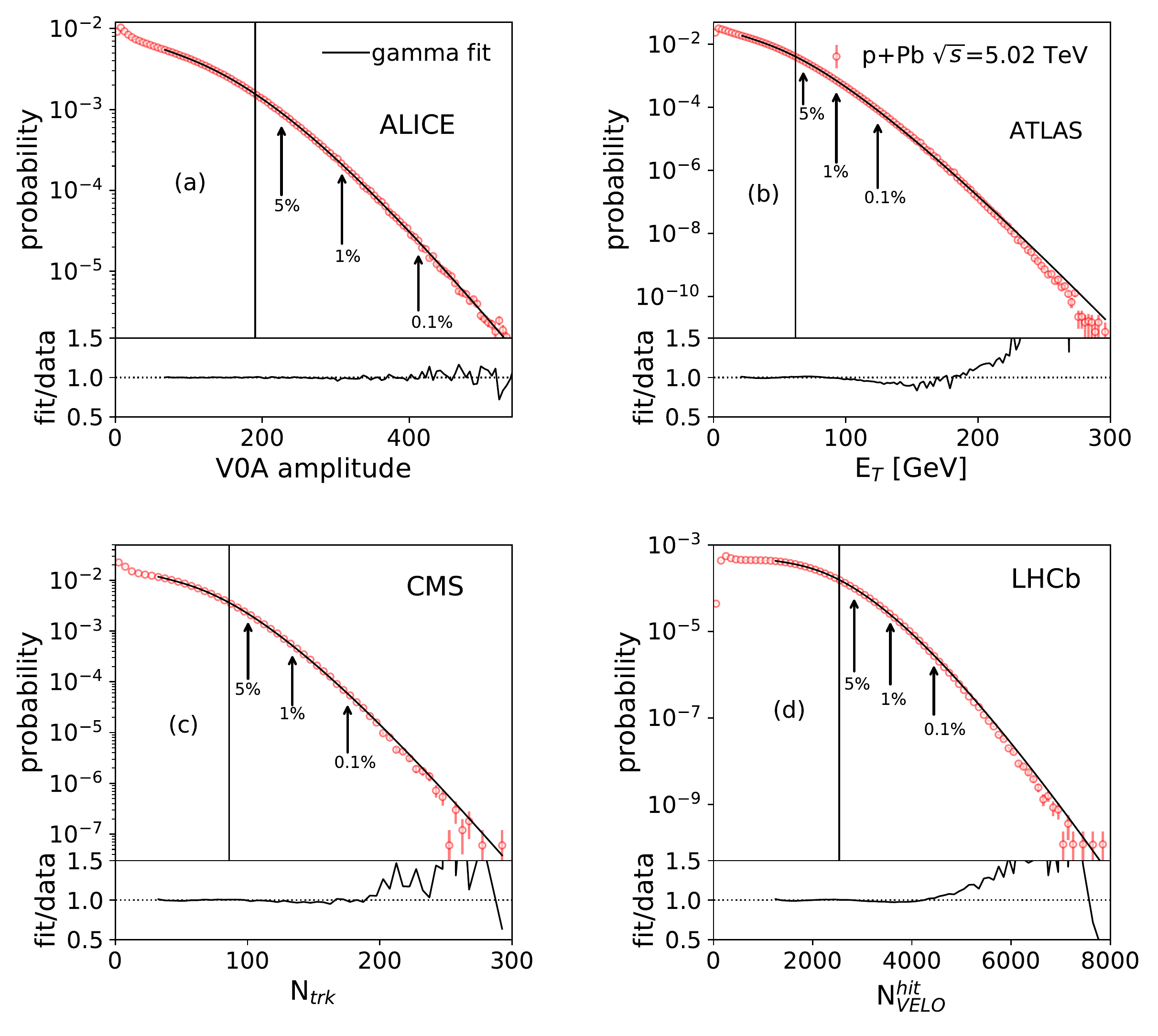} 
\end{center}
\caption{(Color online) 
\label{fig:5}
Distributions $P(n)$ used by all the LHC collaborations to determine the centrality in $p$+Pb collisions at $\sqrt{s_{\rm NN}}=5.02$~TeV. Symbols: experimental data. (a) ALICE data. (b) ATLAS data. (c) CMS data. (d) LHCb data.  Solid lines: fits to the 0-50\% most central collisions using Eqs.~(\ref{cbintegral}), (\ref{paramk}) and (\ref{Gamma2}). Vertical lines in each panel indicate the position of the knee (mean value of $n$ for $b=0$) reconstructed by the fit.
Arrows indicate the values of $n$ corresponding to specific values of the centrality percentile. 
}
\end{figure*} 
We now reconstruct the probability distribution of impact parameter by applying the fitting procedure to LHC data on Pb+Pb collisions at $\sqrt{s_{\rm NN}}=2.76$~TeV, and $p$+Pb collisions at $\sqrt{s_{\rm NN}}=5.02$~TeV.
Table~\ref{tabledata} lists the observables (equivalent of $n$) used to determine the centrality by all the LHC experiments.\footnote{In proton-nucleus collisions, where $n$ is less strongly correlated with impact parameter than in nucleus-nucleus collisions, one sometimes speaks of  ``event activity'' rather than ``centrality''.}
The ALICE collaboration uses the same detectors and observables to characterize the centrality in Pb+Pb and $p$+Pb collisions, and so does the ATLAS collaboration. 
In $p$+Pb collisions, the CMS collaboration uses either the same quantity as in Pb+Pb collisions, namely, the transverse energy in the HF detector~\cite{Chatrchyan:2014hqa}, or the reconstructed track multiplicity~\cite{Chatrchyan:2013nka}. We use only the latter data. As for the LHCb collaboration, data are available only for in $p$+Pb collisions. 

Our procedure is general and applies irrespective of which observable is used to determined the centrality.
The only assumption is that it is an extensive quantity, in the sense that it would be additive if two independent collisions were piled up.
Extensivity can be spoiled if the detector response is nonlinear, which typically happens when the detector is saturated~\cite{Adamczewski-Musch:2017sdk}.
However, our method may still be valid, even in the presence of nonlinearities.
If, for instance, the observed $n$ is a nonlinear function of the incoming multiplicity, but still an increasing function, the ordering of events from less to more central is not modified by the nonlinearity.
Hence, the fraction of events above the knee, which is the key quantity in reconstructing the distribution of impact parameter~\cite{Das:2017ned}, will not be modified.
This can be checked experimentally by carrying out the reconstruction with and without any correction for detector response. 

The distributions $P(n)$ are displayed in Fig.~\ref{fig:5} for the various experiments. 
Results for Pb+Pb collisions were already shown in Ref.~\cite{Das:2017ned},\footnote{The only difference in this paper is that we use a gamma kernel instead of a Gaussian kernel, but the resulting changes in the reconstructed quantities are negligible for Pb+Pb collisions.} so that we only display results for $p$+Pb collisions. 
We fit $P(n)$ using Eqs.~(\ref{cbintegral}), (\ref{paramk}) and (\ref{Gamma2}). For Pb+Pb collisions (not shown), we exclude the most peripheral collisions and only fit the 0--88\% centrality window. As done for the \trento{} data, a polynomial of degree 4 is needed in Eq.~(\ref{paramk}) to achieve a good fit, so that the fit has a total of 6 parameters. For $p$+Pb collisions, as anticipated we fit only the 0--50\%  centrality window, for which a polynomial of order 2 is enough to achieve a good fit, so that the fit has 4 parameters.\footnote{We stress that the results obtained from the fit are very stable under mild variations of the 50\% cutoff.
For all the LHC collaborations, the value of $n_{\rm knee}$ obtained from the fits varies by less than 5\% if the cutoff varies between 30\% and 60\% centrality.}

\begin{figure*}[t!]
\begin{center}
\includegraphics[width=\linewidth]{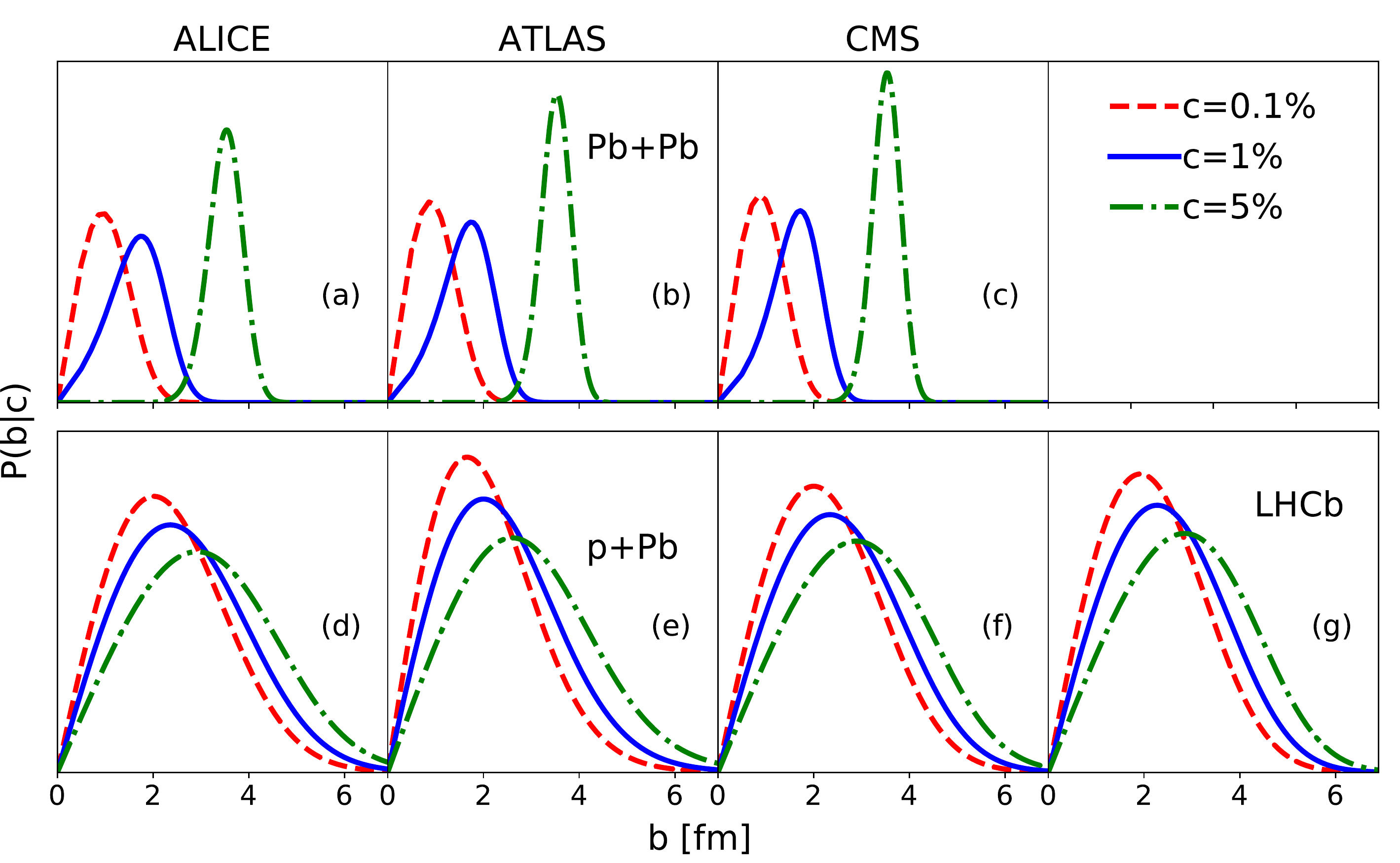} 
\end{center}
\caption{(Color online) 
\label{fig:6}
Distribution of impact parameter reconstructed using Eq.~(\ref{bayes}) for three fixed centralities corresponding to the arrows in Fig.~\ref{fig:5}. (a)--(d): Pb+Pb collisions at $\sqrt{s_{\rm NN}}=2.76$~TeV. (d)--(g): $p$+Pb collisions at $\sqrt{s_{\rm NN}}=5.02$~TeV. (a),(d): using ALICE data.  (b), (e): using ATLAS data. (c),(f): using CMS data. (g): using LHCb data. 
}
\end{figure*} 

\begin{table}[b!]
\begin{tabular}{|c|l|l|}
\hline
&Pb+Pb&$p$+Pb\cr
\hline
ALICE &0.35\%&9.3\%\cr
ATLAS & 0.31\%&7.0\%\cr
CMS &  0.28\% &9.1\%\cr
LHCb &&8.8\%\cr 
\hline
\end{tabular}
\caption{\label{tableknee} 
  Fraction of events above the knee of the distribution for different LHC collaborations in Pb+Pb and $p$+Pb collisions.
  }
\end{table}
We stress that the distribution $P(n)$ is much broader in $p$+Pb than in Pb+Pb collisions.
While the histogram in Pb+Pb collisions has a well-identified tail where $P(n)$ falls off more rapidly (in logarithmic scale), this structure is not clearly visible in $p$+Pb collisions.
This is reflected in the reconstruction by the observation that the fraction of events above the knee, $c_{\rm knee}$, is much larger (see Table~\ref{tableknee}), typically by a factor 30.
This large factor in short explains why the reconstruction of impact parameter turns out to be more difficult in $p$+Pb systems.

Finally, Fig.~\ref{fig:6} displays the distribution of impact parameter reconstructed using Eq.~(\ref{bayes}) for three fixed values of the centrality, for Pb+Pb collisions (top) and $p$+Pb collisions (bottom)\footnote{For ALICE data, we use $\sigma_{\rm inel}=209$~fm$^2$~\cite{Adam:2014qja}. For CMS data, we use $\sigma_{\rm inel}=206$~fm$^2$~\cite{Khachatryan:2015zaa}, and we employ this value as well in the analysis of ATLAS and LHCb data (measurements of $\sigma_{\rm inel}$ in $p$+Pb collisions are not reported by these collaborations).}.
Since for all the LHC collaborations the fits in Fig.~\ref{fig:5} are of excellent quality down to very small probability, the integration in Eq.~(\ref{defc}) that provides the correspondence between $n$ and $c$ is carried out using the fitted $P(n)$, instead of experimental data. 
In other words, we use the fitted $P(n)$ as a smooth interpolation of the histograms, so that we can define $c$ in a continuous range of $n$.  

The distributions shown in Fig.~\ref{fig:6} are very similar for all LHC collaborations, despite the different detectors and acceptances.\footnote{Note, however, that this might change if Relativistic Heavy Ion Collider (RHIC) data were considered, as the corresponding distributions $P(n)$ are much broader than at LHC~\cite{Das:2017ned}.} As shown in Ref.~\cite{Das:2017ned}, the distributions are essentially determined by the centrality of the knee. The values of $c_{\rm knee}$ for all LHC collaborations are given in Table~\ref{tableknee}. The centrality of the knee is similar for all collaborations, which explains why distributions of $b$ are also similar.

Figure~\ref{fig:6}(a)--(c) illustrates that the impact parameter is well reconstructed in Pb+Pb collisions. The distributions are narrow, and there is little overlap between the distributions at $c=1\%$ and $c=5\%$, which means that they correspond to distinct ranges of impact parameter. For centralities in the range $c_{\rm knee}\ll c\ll 1$, $P(b|c)$ is approximately Gaussian, with relative width $\sigma_b/\langle b\rangle\simeq \sqrt{\frac{\pi}{2}}c_{\rm knee}/c$. 
Figures~\ref{fig:6}(d)--(f) show that in p+Pb collisions, on the other hand, the distributions of impact parameter in 1\% and 5\% bins largely overlap, so that the experimental centrality selection is unable to separate events according to impact parameter.
There is however a mild evolution of the impact parameter distribution as the experimental centrality is reduced. In particular, the distribution $P(b|c)$ keeps evolving, although mildly, as one moves towards ``ultracentral'' collisions. 
This must be kept in mind when analyzing ultracentral collisions, either Pb+Pb~\cite{CMS:2013bza,Abelev:2014mda} or p+Pb~\cite{Chatrchyan:2013nka,Abelev:2014mda,Aad:2014lta}.

For sake of simplicity, we have presented distributions of impact parameter at fixed values of the centrality, corresponding to very narrow centrality bins. Extending this reconstruction to a finite centrality bin, corresponding to an interval $n_1<n<n_2$, is straightforward upon integration over $n$:
\begin{eqnarray}
  \label{bayes2}
  P(b|n_1<n<n_2)&=&\frac{\int_{n_1}^{n_2} P(b|n)P(n)dn}{\int_{n_1}^{n_2} P(n)dn}\cr
  &=&P(b)\frac{\int_{n_1}^{n_2} P(n|b)dn}{\int_{n_1}^{n_2} P(n)dn},
\end{eqnarray}
where, in the last equality, we have used Eq.~(\ref{bayes}). Note that
the integral $\int_{n_1}^{n_2} P(n)dn$ appearing in the denominator is simply the width of the centrality bin $\Delta c$ 
(i.e., $0.05$ for the 0-5\% centrality bin). 
Using Eq.~(\ref{Gamma2}), the integral over $n$ in the numerator can be carried out analytically, and one obtains:
\begin{equation}
  \label{bayes3}
  P(b|n_1<n<n_2)=\frac{2\pi b}{\sigma_{\rm inel}\Delta c\,\Gamma(k(c_b))}
\left[\gamma\left(k(c_b),\frac{n}{\theta}\right)\right]_{n_1}^{n_2},
\end{equation}
where $\gamma(k,x)$ denotes the lower incomplete gamma function, and we have used Eq.~(\ref{probab}). 
We apply this equation to ALICE data.
Once we have $P(n)$ from the fit of ALICE data [Fig.~\ref{fig:5}(a)], we insert into Eq.~(\ref{bayes3}) the values of $n_1$ and $n_2$ which correspond to the boundaries of the 0--5\% bin in $c$ (trivially, $n_2=\infty$).
This, along with the reconstructed $k(c_b)$, provides the distribution of impact parameter in the centrality bin, from which we extract $\langle b\rangle=2.85$~fm, and  $\sigma_b=1.37$~fm. By fitting $P(n)$ to a Glauber Monte Carlo model, in the 0--5\% bin the ALICE collaboration obtains $\langle b\rangle=3.12$~fm and $\sigma_b=1.39$~fm~\cite{Adam:2014qja}.
These numbers are consistent with the results displayed in Fig.~\ref{fig:4}(d)--(f): The {\trento} model used in Sec.~\ref{s:validation} is equivalent to a Glauber Monte Carlo, and $\langle b\rangle$ in this model is slightly underestimated by our reconstruction.
However, there is no deep reason to believe that the fit obtained with the Glauber Monte Carlo is more precise than our procedure. The difference between the two simply gives an idea of the uncertainty in our knowledge of the impact parameter. 

\section{Conclusions}

We have provided a method to reconstruct the impact parameter distribution of proton-nucleus and nucleus-nucleus collisions up to $\sim10\%$ centrality.
We use a general Bayesian approach. 
The sole assumption is that the fluctuations of the observable used to determine the centrality at a fixed impact parameter follow a gamma distribution.
In particular, we need not introduce participant nucleons or any measure of the volume of the system~\cite{Zhou:2018fxx}.
The gamma fluctuation kernel is a significant improvement over the Gaussian kernel used previously~\cite{Das:2017ned}, because it has positive support and falls exponentially, thus naturally reproducing the observation in proton-nucleus collisions. 
These improvements come at no additional costs as both distributions have the same number of parameters.
In proton-nucleus collisions, the distribution of impact parameter is broader than in nucleus-nucleus collisions, and depends more mildly on the measured centrality.
Our method does not require any modeling of the collision dynamics, contrary to the estimators of impact parameter currently used in experimental analyses, i.e., Glauber Monte Carlo models.
Therefore, it serves as an independent benchmark in the determination of $b$ from data.


\section*{Acknowledgements}
We thank Alberica Toia and Jiangyong Jia for providing us with
ALICE and ATLAS data.

\end{document}